\documentstyle[12pt]{article}
\begin{document}
\def\theequation{\arabic{section}.\arabic{equation}}
\newcommand{\be}{\begin{equation}}
\newcommand{\ee}{\end{equation}}
\begin{titlepage}
\setcounter{page}{1}
\title{When a mass term does not represent a mass}
\author{V. Faraoni$^1$ and F.I. Cooperstock$^2$ \\ \\
{\small \it $^1$Inter--University Centre for Astronomy and
Astrophysics} \\ 
{\small \it Post Bag 4, Ganeshkhind PO, Pune 411 007, India}\\
{\small \it $^2$Department of Physics and Astronomy, University 
of Victoria} \\ 
{\small \it P.O.
 Box 3055, Victoria, B.C. Canada V8W 3P6}
}
\date{}
\maketitle   
\begin{abstract}
The definition of mass of a scalar field in a curved space has often been
generalized by grouping coupling terms between 
the field and the Ricci curvature with non--curvature--related mass terms. 
In a broader point of view, one sees that a common misunderstanding
resulting from such an identification leads one, in the case of the
spin~2 field, to regard the cosmological constant as a non--vanishing mass
of cosmological origin for the graviton. Similarly, there are
inconsistencies for the spin~1 field. Instead, the intrinsic mass of a
field should be regarded as being independent of the background curvature.
\end{abstract}
\vspace*{1truecm} 
\begin{center}  
To appear in {\em European Journal of Physics}.
\end{center}     
\end{titlepage}   \clearpage \setcounter{page}{2}

\section{Introduction}

The generalization of the Klein--Gordon equation for a scalar 
field $\phi$ on a curved spacetime \cite{footnote1} 
\setcounter{equation}{0}
\be  \label{1}
\Box \phi-m^2 \phi - \xi R \phi =0
\ee
presents the possibility of a non--minimal coupling between the scalar
field and the Ricci curvature $R$ of spacetime. Analogously, direct
couplings with the Ricci and Riemann tensors appear, respectively, in the
wave equations for the electromagnetic field and for gravitational waves
propagating on a curved background spacetime. 
Various authors have grouped the scalar 
field--curvature coupling term $\xi R \phi$ with the
mass term $m^2 \phi$ in Eq.~(\ref{1}), particularly 
when the Ricci curvature of the background spacetime is
constant. In this case, an effective mass $\mu$ given by 
\be   \label{2}
\mu^2=m^2+\xi R 
\ee
is sometimes introduced. Equation~(\ref{2}) 
has led to problems of interpretation. 

The interpretation of the terms coupling the field with the curvature as 
mass terms has also been extended to wave equations for fields of higher
spin.  In fact, Eq.~(\ref{1}) is the prototype of the wave equation on a
curved spacetime, and is  
a guide for the study of more complicated wave equations. For the case 
of spin~2 fields (gravitational waves), the identification of
field--curvature coupling terms with mass terms frequently led 
to attribute 
to the graviton a mass of cosmological origin related to the cosmological 
constant \cite{Tonnelat}--\cite{Curtis}.
Despite some attempts \cite{Treder,TrederYorgrau}, the past and
recent misunderstandings on
this  issue still await clarification in the literature.
The interpretation of terms like $\xi R \phi$ in Eq.~(\ref{1}) as mass
terms leads to
properties of the ``mass'' $\mu$ which are physically unacceptable. 
The proper interpretation is to regard the mass of the field as being
independent of the background curvature.

\vskip2truecm
\section{The Klein--Gordon field}

Intuitively, one would be inclined to consider the mass of a particle as 
an intrinsic characteristic 
which does not depend on whether the particle is in flat space 
or in curved space. The definition~(\ref{2}) of mass of a scalar field 
in curved spacetime does not have this property.

Consider the de Sitter space with arbitrarily large (constant) 
Ricci curvature (i.e. arbitrarily large cosmological constant $\Lambda$),
and
consider Eq.~(\ref{1}) with $m=0$ and $\xi=1/6$ in this space:
\be   \label{4}
\Box \phi-\frac{1}{6}\, R \,\phi=0 \; .
\ee
Equation (\ref{2}) gives a value of the effective mass $\mu$ which is
arbitrarily large. However, the solutions of the wave equation (\ref{4})
propagate on the light cones of de Sitter space. This result is well known
in
the investigations  
of Huygens' principle in curved spacetimes and of ``tails'' of
radiation \cite{Hadamard}--\cite{Gunther}.
The Green function $G(x',x)$ of Eq.~(\ref{1}) is, for a
general given spacetime, the sum of two contributions; the first 
contribution has
support on the light cone, and describes {\em lightlike} propagation of the
waves. The second contribution has support {\em inside} the light cone, and
describes {\em timelike} propagation of waves. In general, the second 
contribution is different from
zero even when $m=0$ in Eq.~(\ref{1}), and is associated with {\em tails} 
(violations of Huygens' principle
\cite{Hadamard}--\cite{Gunther} -- see \cite{FaraoniSonegoPLA} for a
pedagogical introduction).

Equation~(\ref{4}) is conformally invariant and the de
Sitter universe is conformally flat: therefore the second 
contribution to the Green
function 
$G(x',x)$ of Eq.~(\ref{4}) vanishes (no tails) and the
waves propagate strictly on the light cones
\cite{McLenaghan}--\cite{Noonan}.
In other words, 
the tail--free propagation property experienced by scalar waves in flat 
space is transferred to the
conformally flat de Sitter space. Therefore, we have an arbitrarily large
effective mass $\mu=( R/6 )^{1/2}$, but the solutions of Eq.~(\ref{4})
propagate on light cones. It appears reasonable to require that, whatever
definition of mass is adopted in curved spacetimes, particles with nonzero 
mass necessarily propagate strictly {\em inside} the light cone. The 
definition given by
Eq.~(\ref{2}) clearly does not satisfy this requirement.

If $\xi \neq 1/6$, Eq.~(\ref{1})  (with $m=0$) is not conformally
invariant.
It may appear that, in this case, the conclusions reached using our
example with $\xi=1/6$ are not valid and, at most, the previous example
suggests some caution in the
use of Eq.~(\ref{2}). However, the Einstein equivalence
principle \cite{Will} applied to the scalar field $\phi$
restricts the value of $\xi$ to be $1/6$, a result that is now well
established \cite{SonegoFaraoniCQG,GribPoberii,GribRodrigues}.  

The previous example shows clearly the unphysical properties of the mass
$\mu$
defined by Eq.~(\ref{2}). Therefore, the term ``mass'' of the scalar field 
$\phi$ in Eq.~(\ref{1}) should be used exclusively for 
the coefficient $m$, rejecting the
alternative definition (\ref{2}). In other words, the intrinsic mass is
independent of the background curvature. Further support for this
identification is given in the next section. 

\vskip2truecm
\section{The electromagnetic field}

We now consider the Maxwell field tensor $F_{\mu\nu}$ which, in the absence
of sources, satisfies the curved space Maxwell
equations:\setcounter{equation}{0}
\be   \label{maxwell1}
\nabla^{\nu}F_{\mu\nu}=0 \; ,
\ee
\be   
\label{maxwell2}
\nabla _{[\mu}F_{\nu \rho ]}=0 \; .
\ee
Alternatively, the photon field can be described using the four--vector
potential $A^{\mu}$. In the Lorentz gauge $\nabla^{\alpha}A_{\alpha}=0$,
$A^{\mu}$ satisfies the wave equation
\be  \label{5}
\Box A_{\mu}-R_{\mu\nu}A^{\nu}=0   \; .
\ee
We consider again the de Sitter universe, in which the Ricci tensor is
given by
$R_{\rho\sigma}=\Lambda g_{\rho\sigma}$. In this space, Eq.~(\ref{5})
reduces
to 
\be \label{6}
\Box A_{\mu}-\Lambda A_{\mu}=0 \; . 
\ee
According to the identification underlying Eq.~(\ref{2}), 
one would conclude from Eq.~(\ref{6}) that the 
cosmological constant provides a mass for the
photon field. One can actually derive a wave equation for the Maxwell
tensor by
applying the operator $\nabla^{\mu}$ to Eq.~(\ref{maxwell2}) and using
Eq.~(\ref{maxwell1}):
\be    \label{maxwellwave}
\Box F_{\mu\nu}+2R_{\alpha\mu\nu\beta}F^{\beta\alpha}+2R_{\beta [
\mu}{F_{\nu
]}}^{\beta}=0 \; .
\ee
In the de Sitter space one has 
\be \label{RiemanndeSitter}
R_{\mu\nu\rho\sigma}=\frac{2\Lambda}{3} \,g_{\mu [\rho}\, g_{\sigma ] \nu}
\; ,
\ee
which reduces Eq.~(\ref{maxwellwave}) to
\be    \label{AAA}
\Box F_{\mu\nu}-\frac{4\Lambda}{3} \, F^{\mu\nu}=0 \; ,
\ee
in which a ``mass term'' appears, leading to a ``mass'' $\left( 4\Lambda/3
\right)^{1/2} $ for the photon field. This conclusion is incorrect, which
can
be seen as follows: The Maxwell equations (\ref{maxwell1}),
(\ref{maxwell2}) in four dimensions are conformally invariant and the 
de Sitter space is conformally flat. Thus, the electromagnetic field 
propagates
on the light cone \cite{McLenaghan,SonegoFaraoni,Noonan}.
Since the propagation of electromagnetic waves in this case is restricted
to the light cones, we know that the Maxwell field is massless according
to any useful definition of mass for a field in a curved space.

The identification of the linear term in the field as a mass term, as
expressed in Eq.~(\ref{2}) leads to another inconsistency. According to
this 
identification, 
in the case of a spin~1 field in de Sitter spacetime, one would read a
mass $\sqrt{\Lambda}$ from Eq.~(\ref{6}) for the
vector potential $A^{\mu}$. However, the wave equation (\ref{AAA}) for 
the field $F_{\mu\nu}$ would provide a mass $\sqrt{4\Lambda /3}$, so that
the ``mass'' depends on whether one chooses to consider the potential or
the field. This situation is markedly different from the case of the
massive spin~1 field in Minkowski space. 
In flat space, the equations for the Proca 4--potential are 
\be \partial^{\mu}A_{\mu}=0 \; ,
\ee
\be     \label{Proca1}
\Box A^{\mu}-m^2 A^{\mu}=0  \; ,
\ee
where $m$ is the mass of the Proca potential. The Proca field
$F_{\mu\nu}$ satisfies
\be   \label{Proca2}
F_{\mu\nu}\equiv \partial_{\mu} A_{\nu}-\partial_{\nu} A_{\mu} \; ,
\ee
\be                 \label{Proca3}
\partial_{\mu}F^{\mu\nu}=-m^2 A^{\nu}   \; .
\ee
>From Eqs.~(\ref{Proca2}) and (\ref{Proca3}) and from the identity
$\partial_{[\rho}
F_{\mu\nu ]}=0$, it is easy to derive a wave equation for $F_{\mu\nu}$:
\be               \label{Proca4}
\Box F_{\mu\nu}-m^2 F_{\mu\nu}=0  \; .
\ee
Equations (\ref{Proca1}) and (\ref{Proca4}) provide the same value for the
mass of the Proca field, and there is no trace of the
ambiguity encountered in Eqs.~(\ref{6}) and (\ref{AAA}).
          
The presence of terms coupling the field tensor with the Riemann and Ricci
tensors  
follows from the fact that,  in the derivation of the wave equation
for the field tensor $F_{\mu\nu}$ in a curved space, one can take the
combination of covariant derivatives $\nabla_{\mu} \Box A_{\nu}-
\nabla_{\nu} \Box A_{\mu}$ to generate $\Box F_{\mu\nu}$. However, due to
the
non--commutativity of covariant derivatives, contractions of the field
$F_{\mu\nu}$ with the Riemann tensor inevitably appear in the wave
equation for $F_{\mu\nu}$ in addition to the term $\Box F_{\mu\nu}$. By
contrast, in a flat space, $\nabla_{\mu} \Box A_{\nu}-
\nabla_{\nu} \Box A_{\mu}$ is precisely $\Box F_{\mu\nu}$ and hence the 
mass which is
identified from the equation for the 4--vector potential is the same as the
mass that is identified from the equation for $F_{\mu\nu}$.
 
\vskip2truecm                          
\section{Gravitational waves}

   A recurring point in the literature involves the effect of
the cosmological constant in determining the character of gravitational
waves which propagate in curved spacetimes. It has
been explicitly stated or implicitly suggested that the 
cosmological constant endows the graviton with a rest mass
\cite{Tonnelat}--\cite{Curtis}.
This conclusion is incorrect and,
despite some attempts \cite{Treder}, the issue has not been clarified
in the
literature.

The old argument supporting the idea of the cosmological constant as
endowing
the graviton with a mass relies on the fact that the weak field limit of
the 
Einstein equations with a cosmological constant produces a Yukawa (instead
of
a Coulombic) potential \cite{Tonnelat}--\cite{Freundetal}. This
feature has 
been known for many years and its astrophysical consequences have been
explored \cite{Zwicky}. A Yukawa potential is also obtained 
by postulating a finite range of gravitation arising from a nonvanishing
graviton mass \cite{Freundetal}. 

Previous claims \cite{Treder,TrederYorgrau} against the
identification of the cosmological
constant as the mass of the graviton \cite{footnote2} proceed as
follows. An 
exact solution $g^{(0)}_{\mu\nu}$ of the Einstein equations 
with cosmological constant and no matter,\setcounter{equation}{0}
\be                  \label{Einstein}
R_{\mu\nu}-\frac{1}{2} \, g_{\mu\nu}R+\Lambda g_{\mu\nu}=0 
\ee
was considered. This solution was perturbed as
\be
g_{\mu\nu}=g^{(0)}_{\mu\nu}+h_{\mu\nu} \; ,
\ee
where $h_{\mu\nu}$ ($|h_{\mu\nu}|\ll |g^{(0)}_{\mu\nu}|$) describe
gravitational waves. The linearization of the Einstein equations 
(\ref{Einstein}) with the gauge choice
\be   \label{deDonder}
\nabla^{\nu} \left( h_{\mu\nu}-\frac{1}{2} \, g^{(0)}_{\mu\nu} h \right)=0
\ee
(where $h\equiv g^{(0) \mu\nu} h_{\mu\nu}$) gives 
\be          \label{T1}
\Box h_{\mu\nu} -2R_{\alpha\mu\nu\beta}h^{\alpha\beta}
+R_{\mu\rho}{h_{\nu}}^{\rho}+
R_{\nu\rho}{h_{\mu}}^{\rho}-2\Lambda h_{\mu\nu}=0 \; .
\ee
The last term in the left hand side of Eq.~(\ref{T1}) is susceptible 
to being interpreted as a mass term. However, in Ref.~8 it is noted 
that $  R_{\mu\rho}{h_{\nu}}^{\rho}+
R_{\nu\rho}{h_{\mu}}^{\rho}=2\Lambda h_{\mu\nu}$. Using this result, 
Eq.~(\ref{T1}) is reduced to
\be   \label{T2}
\Box h_{\mu\nu} -2R_{\alpha
\mu\nu\beta}h^{\alpha\beta}=0 \;.
\ee 
According to Ref.~8, the substitution has 
removed all trace of $\Lambda $ and 
the propagation equation is reduced to the same form as it would have in
a spacetime free of any cosmological constant. The claim that this proves
that the propagation describes a massless field on this basis is
unfounded: 
if one proceeds to substitute the form~(\ref{RiemanndeSitter}) of the
Riemann tensor, one obtains
\be   \label{massagain}
\Box h_{\mu\nu} -\frac{2\Lambda}{3} \, h_{\mu\nu}=0 \;,
\ee 
in which a ``mass term'' re--appears in the wave equation for
gravitational 
waves, and one is presented with a quandary. Equation (\ref{massagain})
agrees with Eq.~(2.21) of Ref.~21 and is obtained by imposing
(\ref{deDonder}) and the additional constraint $h=0$ (the propagation 
equations for $\nabla^{\nu}h_{\mu\nu}$, $h$ and a proof that the
constraints
$\nabla^{\nu}h_{\mu\nu}=0$, $h=0$ can be imposed in a globally hyperbolic 
spacetime can be found, e.g., in Ref. \cite{Higuchi}).

Armed with our argument from the previous 
section, we recognize that the identification of the second 
term in the left hand
side of Eq.~(\ref{massagain}) as a mass term is inappropriate. Thus, it is
seen
that the correct conclusion was reached in Ref.~8, but the proof was 
incomplete. While the presence of $\Lambda$ affects the propagation of
scalar, vector and tensor fields in curved spacetime, it does not endow the
fields with intrinsic mass.

The connection between the cosmological constant and the mass of the
graviton
is excluded. Although
the introduction of the quantity $\mu$ in Eq.~(\ref{2}) may be useful from
the 
mathematical point of view in some cases, it should not 
be identified with the physical mass of the field, since such an 
identification would lead to unacceptable physical properties for the
field. 

\vskip2truecm
\section*{Acknowledgment}

This work was supported, in part, by the Natural Sciences and Engineering
Research Council of Canada. 

\clearpage          

{\small  
\begin{thebibliography}{99}
\bibitem{footnote1} The metric signature is --~+~+~+. Greek indices
assume the values 0,~1,~2,~3, Latin indices assume the values 1,~2,~3. The
Riemann tensor
is given in terms of the Christoffel symbols by
${R_{\mu\nu\rho}}^{\sigma}=
\Gamma^{\sigma}_{\mu\rho ,\nu}-\Gamma^{\sigma}_{\nu\rho ,\mu}+
\Gamma^{\alpha}_{\mu\rho}\Gamma^{\sigma}_{\alpha\nu}-
\Gamma^{\alpha}_{\nu\rho}\Gamma^{\sigma}_{\alpha\mu} $. The Ricci tensor
is $R_{\mu\rho}\equiv {R_{\mu\nu\rho}}^{\nu}$. Round [square] brackets
denote [anti]symmetrization. $\nabla_{\mu}$ is the covariant derivative
operator, and
$\Box \equiv g^{\mu\nu}\nabla_{\mu}\nabla_{\nu}$.
\bibitem{footnote2} In Ref.~\cite{Treder}, 
a second argument is presented, different from the one that we consider 
in this paper. The second argument relies on the perturbation of a flat 
background, which is unacceptable since the flat metric that is perturbed 
is {\em not} a solution of Eq.~(\ref{Einstein}) if $\Lambda \neq 0$.
\end{thebibliography}
\end{document}